\def\noi{\noindent}
\def\pref#1{(\ref{#1})}    
\def\beq{\begin{equation}}
\def\eeq{\end{equation}}
\def\bmulteq{\begin{eqnarray}}
\def\emulteq{\end{eqnarray}}

\def\SM{Standard Model}
\def\MSSM{Minimal Supersymmetric Standard Model}

\def\CPV{$CP$ violation}

\def\susy{supersymmetry}

\def\susyic{supersymmetric}
\def\Susyic{Supersymmetric}

\def\dn{$d_n$}
\def\de{$d_e$}

\def\ecm{e$\,$cm}
\def\ecmeq{{\rm e\, cm}}

\def \Im{{\rm Im}}

\def \Arg{{\rm Arg}}

\def\comment{\setbox99=\vbox}

\def\refmark{\cite}
\def\Ref{\bibitem}
\def\eqn{\label}

\def\singlespace{\renewcommand{\baselinestretch}{1} \small\normalsize}
\def\oneandahalfspace{\renewcommand{\baselinestretch}{1.5} \small\normalsize}
\def\footnt#1{$\!\!\!\!$ {\singlespace \footnote{#1}\oneandahalfspace}}

\def\sAg{$\sin\phi_{A\lambda}$}
\def\smug{$\sin\phi_{\mu \lambda}$}

\def\Kpot{K\" ahler potential}

\def\Arg{\mbox{\rm Arg}}

\def\abs #1{|#1|}

\def\abs #1{|#1|}

\def\H{\mbox{\rm H}}

\def\FullA    {A^* m_0 - \mu \, v_u/v_d }

\def\plet #1 {\left\lgroup \matrix{#1} \right\rgroup}

\def \vev#1{\left< #1 \right>}

\documentstyle[12pt]{article}

\renewcommand{\baselinestretch}{1.5}

\begin{document}

\def\preprint#1{{\vskip -.25in\hfill\hfil#1\vskip 10pt}}

\vskip -1 in
\preprint{TRI-PP-24}
\preprint{May 1993}

\centerline{{\Large\bf Natural CP Violation Criteria}}
\centerline{{\Large\bf for the Minimal Supersymmetric Standard Model}}
\centerline{Robert Garisto}
\begin{center}
{\it TRIUMF,
4004 Wesbrook Mall,
Vancouver, B.C.,
V6T 2A3,
Canada}
\end{center}

\begin{abstract}
$CP$ violating effects in the Minimal Supersymmetric Standard Model (MSSM)
can lead to excessively large contributions to the neutron EDM (\dn).
We write criteria which ensure that
the low energy supergravity (SUGRA) parametrization of
the MSSM does not require fine-tunings or large mass scales to
evade the constraint from \dn, and consider the implications on
SUGRA theories.
In particular, we show that in the Polonyi model,
two of the mass scales are in general complex, meaning
that model does not naturally avoid a large \dn\ as is sometimes claimed.
\end{abstract}


\section{Introduction}

\Susyic\ (SUSY) theories are perhaps the most widely considered extensions to
the \SM\ (SM)
\cite{REFa}.
This is in large measure due to the fact that they are
the only known perturbative solutions to the naturalness problem
\refmark{Ibanez}. Since this means SUSY theories are attractive partially
because they  remove  fine-tunings present in the SM
(in quadratically divergent radiative  corrections  to the
Higgs mass), we believe that new fine-tunings should not be introduced
to satisfy other phenomenological constraints.  We also
would like to be able to break $SU_2 \times U_1$ radiatively.
Both of these considerations lead one to conclude that large
superpartner masses are disfavored
\cite{REFay,private Kane}.

It is well known
\comment{
\cite{historical,Fischler et al,Kizukuri et al,REFc,REFam}
\cite{REFbe,REFj,REFbd,Dugan et al,REFab,REFbc,Gilson}
}
\cite{historical}-\cite{Gilson}
that for moderate mass scales, SUSY predicts a neutron electric dipole
moment (\dn) of order $10^{-{\rm 22-23}} \tilde\varphi$ \ecm, where
$\tilde \varphi$ is some combination of SUSY phases.
Since the experimental upper bound on \dn\ is extremely small, now below
$10^{-25}$\ecm\ \cite{Ramsey talk},
this constitutes a fine-tuning problem.
One would like a set of criteria which ensures that a given model avoids
this `SUSY \dn\ problem' naturally.
Our approach is to write these criteria in a way which is useful
for supergravity (SUGRA) model building.
Using our criteria, it will be easy to show that the Polonyi model
of SUGRA does not naturally avoid a large \dn\ as is sometimes claimed.


We write the superpotential of the \MSSM\ (MSSM) as
\beq
W = W_Y + \mu \H_u \times \H_d,
\label{MSSM superpot}
\eeq

\noi
where $\H_u$  and $\H_d$ are Higgs doublet superfields,
and $W_Y$ contains the Yukawa sector of the theory.
The Higgs mixing coefficient $\mu$ can be complex.
There are also complex phases in the soft breaking potential
( ${\cal L}_{soft}$). We will use the low energy supergravity parametrization
of ${\cal L}_{soft}$ \cite{REFh}:
\bmulteq
-{\cal L}_{soft} &&= \abs{m_i}^2 \abs{\varphi_i}^2 + \nonumber\\
&&\left( {1 \over2} \sum_\lambda\tilde m_\lambda   \lambda \lambda  +
       A m_0^* \left[ W_Y \right]_\varphi  +
       B m_0^* \left[ \mu \H_u \times \H_d \right]_\varphi  + h.c.\right).
\eqn{LsoftSUGRA}
\emulteq

\noi
where $\varphi_i$ are the scalar superpartners, $\lambda$ are the
gauginos, and $[\ ]_\varphi$ means take the scalar part.
The parameters $A$ and $B$, their associated mass scale $m_0$,
and
the common gaugino mass $\tilde m_\lambda$,
are in general {\it complex},
and thus contribute to $CP$ violating effects.
The phase of $m_0$ is
often overlooked.
We could lump $m_0$ into the definitions
of $A$ and $B$, by defining $\bar A \equiv A m_0^*$
and $\bar B \equiv B m_0^*$,
but that obscures the fact that $A$ and $m_0$ come from different
places in the SUGRA theory, while $A$ and $B$ are simply related
by $B=A-1$
\refmark{REFe}.
As a dramatic illustration of this point, we
will show in Section 4 
that the Polonyi model of SUGRA
gives $A$ and $B$ real, but $m_0$ is complex,
so that
$\bar A$ and $\bar B$ are both complex and thus
contribute to \dn.

The largest SUSY contributions to \dn\ tend to come from squark mixing in
gluino-squark loops, as in Figure 1.
The LR mixing pieces of the down squark mass matrix can be written as
$M^2_{LR} = (A^* m_0 - \mu \, v_u/v_d)\, \hat M_D$, where $v_{u,\, d}$ are
the Higgs vacuum expectation values (VEVs), and $\hat M_D$ is the diagonal down
$quark$ mass matrix.  Note that several
key references omit the $\mu$ term
\refmark{REFbe,REFab,REFe,REFac,REFt}.
Figure 1 gives
\cite{REFab}

\beq
d_n(\tilde g) =
\left({8\, e \, \alpha_s \over 27 \pi}\>
I\left({\tilde m_g^2 \over \tilde m_d^2} \right) \right)\,
{\,  \Im\left[ \left( \FullA\ \right) \, \tilde m_g\right] \,
m_d \, \over {\tilde m_d}^4}
, \eqn{dnintofImA}
\eeq

\noi
where $\tilde m_g$ is the gluino mass, $m_d$ is the current down quark
mass, and $\tilde m_d$ is the average down $squark$ mass.
The integral $I(X)$ \cite{Kl3},

\beq
I(X) ={1\over (1-X)^2} \left[{1\over 2} (1+X) + {X\over 1-X}\ln X \right]\ ,
\eqn{AIb}
\eeq

\noi
appears incorrectly in \cite{REFab} but is corrected in \cite{REFac}.
Note that some papers which cite \cite{REFab} quote the
incorrect integral
\refmark{REFbe,REFj}.   
Also note that \refmark{REFbe,REFab}
write $d_n$ in a way which gives the appearance
that \dn\ is proportional to the mass squared
difference between the squarks.
Their expressions are correct, but misleading, because the mixing
angle between the squarks goes as the inverse of that mass difference,
leaving \dn\ of the form given above.


The two phases which come into \pref{dnintofImA}  are
$\varphi_{A\lambda} \equiv \Arg (A^* \, m_0 \, {\tilde m_\lambda}),$ and
$\varphi_{\mu \lambda} \equiv \Arg (\mu \, v_u/v_d \> \tilde m_\lambda)$.
We have used the fact that the phase of the gluino mass is that
of the common gaugino mass $\tilde m_\lambda$.
We define a new SUSY mass scale,
$\tilde M^2 \equiv \tilde m_d^4 / |m_0| \, \tilde m_g$,
so as to group all of the SUSY mass scale behavior in one place.
We expect this $\tilde M$  to be of order
the weak scale, because larger values can cause fine-tuning problems.
For example, Ross and Roberts
\cite{REFay}
find they run into fine-tuning problems
if the SUSY mass scale is greater than about $3M_Z$.
Finally, we take $\alpha_s=0.12,   
m_d =$ 10 MeV,
define $\tan{\beta} \equiv \abs{v_u/v_d}$,
and normalize $\tilde M$ to 100 GeV to obtain

\beq
d_n \simeq 5 \times 10^{-23} \ecmeq\ {\left( {100 \mbox{GeV} \over \tilde M}
\right)}^2 \left[ \abs{A} \sin {\varphi_{A\lambda}} - {\abs{\mu} \over
|m_0|} \tan{ \beta} \sin{\varphi_{\mu \lambda}} \right] \ .
\eqn{seconddnOOM}
\eeq

\noi
There are other contributions\footnt{There will also be strong
CPV effects, but these occur in the
SM as well and presumably  a solution to the strong $CP$ problem
will not change our conclusions about weak CPV.}
to \dn\ from up squark,
chargino and neutralino mixing, which have different combinations
of these phases ($e.g.$ the up squark contribution goes as $\cot\beta$ instead
of $\tan\beta$), so that a cancellation between the separate pieces would
be a fine-tuning.

It is clear there are three ways we can make \pref{seconddnOOM} satisfy the
experimental bound on \dn,
which is now below $10^{-25}$\ecm\ \cite{Ramsey talk}:

\

\setbox5 = \vbox{\hsize=5 in \leftskip= .5 in
\baselineskip= 10pt

\noi (i) fine-tune the phases (individually or in combination)
to order $10^{-2}$--$10^{-3}$.

\

\noi (ii) require a large scale $\tilde M$, of order a few TeV.

\

\noi (iii) restrict ourselves to models in which
\sAg\ and \smug\ are naturally zero.

}

\box5

\noi
As we said, (i) is simply unacceptable.
Several authors use (ii)
\refmark{Kizukuri et al,REFc,REFl}.
In addition to the
fine-tuning problems mentioned above, such models
may also cause cosmological problems: if all the sfermion
masses exceed about 400 GeV, the lightest \susyic\ partner (LSP)
annihilation cross section
will be too small, leaving an LSP relic density with $\Omega>1$
\cite{Leszek DM review}.
We have also examined the data from models of
Kane, Kolda, Roszkowski and Wells
\cite{Kane etal} which satisfy all other
important phenomenological constraints.
We find that fine-tunings between $10^{-2}$ to $10^{-3}$ are required
in almost all of these models.
So we choose to explore option (iii), and develop criteria
which ensure that
\sAg\ and \smug\ are naturally zero.


\section{Phases Criteria}

Let us henceforth assume that we are working in a MSSM with a scale
$\tilde M$ of order the weak scale, and that we do not accept fine-tunings
as a solution to the SUSY \dn\ problem.  These assumptions lead us to
conclude that our model must satisfy the criteria in (iii) of the
last section, $i.e.$ we {\it need}
the two physical phases to be zero, $i.e.$
$\sin\phi_{A\lambda}$ and $\sin\phi_{\mu \lambda} =0$.
Setting these phases to zero has been discussed in several references;
the purpose of this section is to rewrite these criteria
in a way most useful to model builders. We will then be able
to explore the implications of these criteria on SUGRA models.
A slightly different approach was taken by Kurimoto \cite{Kurimoto},
which is equivalent to the first part of our discussion.

The first thing we can do is rotate away the phase of $\mu$.
The $\mu$ dependence comes from two sources: the soft breaking term in
\pref{LsoftSUGRA}, which is responsible for
the $\mu_{12}^2$ Higgs mixing term in the scalar potential
\cite{REFg},
and the $F$ term, which contributes to everything else.
We can rotate the relative phase of the Higgs superfields $\H_u$ and $\H_d$
 so that the $\mu_{12}^2$ term is real.
The Higgs potential will be real, so that $v_u/v_d$ will be real,
the phase of $\mu$ in all the $F$ terms will be changed to that
of $B^* m_0$, and the Yukawa couplings will absorb the phase after a
quark field rotation.  Thus there is no remaining trace of the
original phase of $\mu$.

Crucial to this procedure is the assumption that the $\mu_{12}^2$ term
comes from the superpotential.  If the soft breaking scalar potential
Higgs mixing term were put in by hand, then $\mu_{12}^2$ would be unrelated to
$\mu$, and the phase of $\mu$ would contribute to \CPV.
We discuss this below.

At this point, we have established that one can write the squark
mixing contribution to \dn\ in terms of
$\Arg(A^* m_0 \tilde m_{\lambda})$ and $\Arg(B^* m_0 \tilde m_{\lambda})$.
In SUGRA theories, one often  obtains $B = A -1$.
{\it If} $B=A-1$
holds,\footnt{If this relation does not hold (as with an effective $B$
described
below), one simply adds the condition that $B$ be real to \pref{crit three}
while \pref{crit four} is unchanged.}
then one can write our criteria as:

\beq
A\ \&\ (m_0 \tilde m_{\lambda}) \ \mbox{\rm must be real}.
\label{crit three}
\eeq

\noi
These criteria should be satisfied at the high energy scale.
The parameters will remain real as they evolve down to the weak scale.

There is one more degree of freedom we can rotate: the phase of the
Grassmann variable $\theta$.\footnt{This is equivalent to an
R-rotation with all superfields having
R-character zero.  It does
{\it not} assume anything about the R-invariance of the Lagrangian.}
This freedom allows us to rotate away
the phase of $m_0^* \tilde m_\lambda$, which means that
if $m_0 \tilde m_\lambda$ is real, we can always arrange to have
$m_0$ and $\tilde m_\lambda$ separately real.
In addition, we know that the phase of $\mu$ can be rotated away.
Thus $if$ our criteria of \pref{crit three} are satisfied,
we can always write the theory such that

\beq
A,\ B,\ m_0,\ \tilde m_{\lambda},\ \&\ \mu \ \mbox{\rm are real}.
\label{crit four}
\eeq

\noi
It is clear
that our criteria in \pref{crit three}
are sufficient because \pref{crit four} says that
all SUSY parameters are real---there is no new SUSY contribution to \dn.

Important special cases are those
in which one or more of the parameters are zero.
Then one must worry about other phenomenological constraints---for example
if $\mu,\ B$ or $m_0$ were zero, then
 $\mu_{12}^2$ would be zero and the theory
would develop an unacceptable massless axion \cite{KIMaNILLES}.
To avoid this, one could relax
the assumption that the bilinear soft breaking term comes from the
superpotential, and put $\mu_{12}^2$ in by hand, though this may not
fit into a SUGRA derived theory.
Thus there are two cases to consider:

\setbox6 = \vbox{\hsize=5 in \leftskip= .5 in
\baselineskip= 10pt

\noi
Case I:  $\mu_{12}^2 = B m_0^* \mu$, $i.e.$ the soft breaking
Higgs mixing term comes from the Higgs mixing term in the superpotential.

\

\noi
Case II:  $\mu_{12}^2$ is not related to $\mu$, $e.g.$ the soft breaking
terms are put in by hand.

}

\

\box6

\noi
In case II, the phase of $\mu$ cannot be rotated away, but we can
lump its phase into an effective parameter
$\bar B_{eff} \equiv \mu_{12}^2 /\mu$
(which one identifies with $\bar B\equiv B m_0^*$ of case I), and proceed
as if we were in case I, keeping $\bar A\equiv A m_0^*$ and $\bar B_{eff}$
as separate parameters.\footnt{We note in passing
that the {\it sign} of $\mu$ {\it before}
rotation is not physical.
After rotation of the Higgs superfields, the sign of the
Higgs mixing $F$-terms is that of $\bar B_{eff}$ from {\it before} the
rotation.
For example, a SUGRA model (in case I)
which gives a $B>0$ ($B<0$), will after
rotation give a positive (negative) coefficient $\mu$ (in the basis of
positive $\mu_{12}^2$), regardless of the
original sign of $\mu$.}
The criteria in case II are best written:

\beq
\bar A,\ \bar B_{eff},\ \&\ \tilde m_{\lambda}\ \mbox{\rm must be real}.
\label{crit five}
\eeq

\noi
Of course the effective $B$ prescription fails if $\mu=0$,
but that case yields a massless higgsino \cite{REFa}
and is thus ruled out.
The criteria in \pref{crit five} also imply that we can write the theory such
that \pref{crit four} holds (with $B$
replaced by $B_{eff}\equiv \bar B_{eff}/m_0$), $i.e.$ there is no new
SUSY $CP$ violation.

The only remaining possibilities (in either case I or case II)
are for $\bar A$ and/or $\tilde m_\lambda$ to be zero.
Having $\tilde m_\lambda = 0$ allows one to rotate away one phase, leaving
one physical SUSY phase. For example, if $B=A-1$, the only requirement to
avoid a large \dn\ would be that $A$ is real, which is satisfied in
some SUGRA models.  The case $\tilde m_\lambda = 0, \bar A=0$ has {\it no}
 physical SUSY phases (we can rotate away the phase of $\bar B_{eff}$),
and thus would also solve the SUSY \dn\ problem.
Unfortunately for both of these solutions,
$\tilde m_\lambda = 0$ means that the gluino is massless
at tree level, leaving the gluino with a loop generated mass
which is far too small.\footnt{Recently there has been
a revival of the concept of a `light
gluino window' \cite{light gluino}, but this
possibility is almost certainly ruled out experimentally
\cite{Kane gluino}.
}

The remaining possibility, $\bar A=0$ (but $\tilde m_\lambda \neq 0$),
offers no improvement over the criteria in  \pref{crit five}, for one
still has to make $\bar B_{eff}$ and $\tilde m_\lambda$ real.

In summary, there are no phenomenologically viable solutions to the SUSY \dn\
problem  obtained
from setting any of the parameters zero which are not included explicitly in
\pref{crit three} or \pref{crit five}, and each of these imply that there
is no new SUSY \CPV\ in the MSSM.
Of course model builders may find it easier
to make a parameter zero rather than simply real, but one must be careful
that no other phenomenological constraints are violated.


\section{Radiative Effects}

We must be sure that radiative effects do not change our conclusions.
If the parameters are initially real,
they will remain so when evolved to low energy
\cite{RGE 1,RGE 2,RGE 3}.
There are CKM dependent terms induced into the squark mass matrix, but
these were found to give \dn\
of order $10^{-31}$\ecm \cite{Dugan et al}.

However, this does not include
possible $finite$ effects to the squark LR mixing of the form

\beq
\delta M_{LR}^2 = k \bar A V_L^\dagger X V_L \hat M_D,
\label{MLR}
\eeq

\noi
where $X$ can be off-diagonal.
The contribution to \pref{MLR} from Figure 2 gives

\beq
X= \sum_{n=1}^2 \Omega_{Ln}^\dagger I(n) \Omega_{Ln} ,
\eeq

\noi
where $I(n)$ is a diagonal $2 N_F \times 2 N_F$ matrix of integrals, and

\beq
U_{II}^\dagger \equiv
       \pmatrix{\Omega_{L1} &\Omega_{R1}\cr
                \Omega_{L2} &\Omega_{R2}\cr}
\label{defOmega}
\eeq

\noi
diagonalizes the up squark mass matrix.
The $3\times3$
matrix $\Omega_{L1}$ has large off-diagonal parts, but
the off-diagonal parts of $X$ turn out to be proportional to squark
mass splittings, and the contribution to \dn\ from these finite
pieces is negligible, of order $10^{-37}$\ecm.
Finite CKM corrections to the
gluino mass were addressed in \cite{REFbc} and also found to be small.

We note that one
could have {\it spontaneous} \CPV\ induced
through loop effects to the Higgs potential
in the MSSM \cite{Maekawa},  though this possibility has been ruled out because
it gives a $CP$ odd Higgs scalar which is too
light \cite{Pomarol}. So we conclude that radiative effects do not
affect our criteria.


\section{Supergravity Criteria}

We would like to see how SUGRA theories
(for a review see for example \cite{REFh,REFe,REFt,REFf})
fare with respect to the SUSY \dn\ problem.
We consider models of the superhiggs effect, where the VEVs of
a hidden sector break \susy, and
the soft SUSY breaking parameters are determined by inputs
to the underlying SUGRA
theory.  Using our criteria from Section 2, we derive criteria for
prospective SUGRA models
which would naturally solve the SUSY \dn\ problem.
We then  examine the argument that
a particular SUGRA model, the Polonyi model
\cite{REFpolonyi},
provides such a solution.
While it turns out that the
soft trilinear coefficient $A$ is {\it real}
\refmark{REFbe,REFj,REFh,REFe,REFt,REFl,REFu,REFy},
the argument that this is a solution
\refmark{REFbe,REFl}
to the SUSY \dn\ problem is incorrect, since the Polonyi model
cannot in general require $m_0$ or $\tilde m_\lambda$ to be real.

\subsection{The Superhiggs Effect}

We begin with the SUGRA scalar potential
\refmark{REFx}

\beq
V= e^{K/M^2} \left[ \left(D_i w \right) \left(D_{j^*} w^* \right) g^{i j^*}
- 3 {\abs{w}^2 \over M^2} \right]
+ {1\over 2} f^{-1}_{\alpha \beta} D^\alpha D^\beta
, \eqn{SUGVDw}
\eeq

\noi
where $M$ is $(8 \pi)^{-1/2}$ times the Planck mass, $w$ is the superpotential,
$D^\alpha$ are the auxiliary fields,
$K$ is the \Kpot,
and $g^{i j^*} = \partial^2 K / \partial\phi_i \partial\phi_j^*$.
We define the K\" ahler derivative, $D_i$, as
\beq
D_i w \equiv {\partial w \over \partial \varphi_i} +
    {1 \over M^2} {\partial K \over \partial \varphi_i} w .
\eqn{Kcovderiv}
\eeq

\noi
SUSY is broken by $\vev{D_i w}$, giving a common gaugino mass
\refmark{REFh}

\beq
\tilde m_\lambda = e^{<K>/2M^2} \left< g^{ij^*}\right>
\vev{ f_{,i}} \left< D_{j^*} w^*\right> ,
\eqn{gauginomass}
\eeq

\noi
where $f_{,i}\equiv \partial f(\varphi^i)/\partial \varphi^i$ and
we have assumed that the gauge kinetic metric $f_{\alpha\beta}$
is diagonal in its gauge indices, $i.e.$ $f_{\alpha\beta}(\varphi_i) =
f(\varphi_i) \delta_{\alpha\beta}$.
The gauginos are massless at tree level unless $f$ is a non-trivial
function of $\varphi_i$.

To analyze the $CP$ violating effects of SUGRA models, we must
find the low energy scalar potential.
We make the usual assumption that $w$
can be divided into a sum of a {\it hidden sector} and a {\it visible
sector}
\cite{REFe},   
\beq
w(z_i, y_a) = h(z_i) + g(y_a) ,\eqn{Whplusg}
\eeq

\noi
where $y_a$ are {\it visible} scalar fields, which interact
with \SM\ particles, and $z_i$ are {\it hidden} scalar fields,
which interact with \SM\ particles only through gravity.  The
VEVs of the $z_i$ break SUSY and give a mass to the gravitino
and to the scalars in $V_{soft}$. Plugging \pref{Whplusg} into
\pref{SUGVDw} and using the definition  \pref{Kcovderiv}, we find

\bmulteq
V= e^{K/M^2} \Biggl[
\left(h_{,i} + { h + g\over M^2} K_{,i} \right)
\left(h^*_{,j^*} + { h^* + g^*\over M^2} K^*_{,j^*} \right)
g^{i j^*}&\nonumber\\
+
\left| g_{,a} + { h + g\over M^2} y_{a}\right|^2
- 3 {\abs{h+g}^2 \over M^2} \Biggr]
+ {1\over 2} f^{-1}_{\alpha \beta} D^\alpha D^\beta .& \eqn{SUGVhg}
\emulteq

Now we want  to break SUSY by letting $z_i$ have a VEV of order $M$
\cite{REFe},

\beq
\vev{z_i} = b^0_i\, M ,\
\vev{K} = \tilde b^2 \, M^2 , \ \vev{K_{,i}} = b_i^* \, M ,
\eqn{defKVEV}
\eeq

\noi
where $b^0_i$ is some {\it complex} constant of order one.
We have defined $\tilde b^2$ and $b_i$ such that
in the flat case (where $K_{,i} = z_i^*$),
$\tilde b^2 \rightarrow \abs{b^0_i}^2$ and $b_i \rightarrow b^0_i$.
We need a hidden sector potential, $h(z_i)$, which contains a small scale $m$:
\beq
\vev{h} = m \, M^2 ,\
\vev{h_{,i}} = a^*_i \, m \, M , \eqn{hpVEVdef}
\eeq

\noi
where $a_i$ is some other  complex constant of order one.
Using \pref{defKVEV} and \pref{hpVEVdef} in \pref{SUGVhg},
the condition for no Planck scale
cosmological constant is

\beq
(a_i^* + b_i^*)(a_{j^*} + b_{j^*}) \vev{g^{ij^*}} = 3 , \eqn{cosmocondfull}
\eeq

\noi
which for the flat case just gives the usual
\refmark{REFe}
expression $ \abs{a_i + b_i}^2 = 3$.

Let us define the low energy superpotential
$[W]_\varphi \equiv e ^{{1\over 2}\tilde b^2} g$,
and mass

\beq
m_0 \equiv e ^{{1\over 2}\tilde b^{2*}} m
= e ^{{1\over 2}\tilde b^{2*}}  {\vev{h} \over M^2} , \eqn{modef}
\eeq

\noi
which in the flat case has a magnitude of the gravitino mass, $m_{3/2}$.
We further define the parameter

\beq
A \equiv b_i^*(a_{j^*} + b_{j^*}) \vev{g^{ij^*}} , \eqn{Anonmin}
\eeq

\noi
which is a generalized definition of the usual
flat K\" ahler metric definition
\refmark{REFe}
$A \rightarrow b_i^{0\, *}(a_i + b^0_i)$.
Putting these definitions into \pref{SUGVhg}, we obtain
the low energy supergravity scalar potential:

\bmulteq
&&V(\varphi_i) = | F_i |^2 + {1\over 2}D_a D^a + V_{soft},\nonumber\\
&&V_{soft} = \abs{m_0}^2 \abs{\varphi_i}^2 +
\left( A m_0^* \left[ W^{(3)}\right]_\varphi  +
       B m_0^* \left[ W^{(2)}\right]_\varphi  + h.c.\right),
\eqn{VofAB}
\emulteq

\noi
where the parameter $B=A-1$.
Here $W^{(2)}$ and $W^{(3)}$ are the quadratic and cubic terms in
the superpotential.

Now we can write the criteria from Section 2
in terms of SUGRA parameters.
Since $\mu_{12}^2 = B m_0^*\mu$ (Case I), and $B=A-1$,
we can use the criteria in \pref{crit three}.  The
first criterion, that $A$ must be real, is satisfied by \pref{Anonmin}
if $a_i$ and $b_i$ are {\it relatively} real, and if $\vev{g^{i j*}}$
is real.  The latter is true in the flat case.  We will see that
the former is true at least in the Polonyi model.

The other criterion of \pref{crit three} is that $(m_0 \tilde m_\lambda)$
must be real.  It turns out that $\tilde m_\lambda$ in \pref{gauginomass}
can be written proportional to $m_0^*$, so that

\beq
m_0 \tilde m_\lambda = \abs{m_0}^2 \vev{M f_{,i}}
\left(a_{j^*} + b_{j^*}\right) \vev{g^{i j^*}},
\label{m0 mlambda}
\eeq

\noi
which is real if the coefficients $a_i$ and $b_i$ are {\it real} (not just
relatively real), and if
$\vev{g^{i j^*}}$ and $\vev{f_{,i}}$ are real.  These contain the
criteria which make $A$ real.
Thus our criteria \pref{crit three} can be written:

\beq
a_i,\ b_i,\ \vev{f_{,i}}, \vev{g^{ij^*}} \ \mbox{\rm must be real},
\label{crit seven}
\eeq

\noi
where $a_i$ and $b_i$ are defined by \pref{hpVEVdef} and \pref{defKVEV}
respectively.
If, for example, $f$ can be
written in the simple form $f(z_i) = c_n z_i^n/M^n$, then the only phases
in $\vev{f_{,i}}$ will be those of $c_n$ and $b_i^0$, so that a
sufficient condition for solving the SUSY \dn\ problem in the
flat case would be:

\beq
a_i,\ b_i,\ c_n  \ \mbox{\rm must be real}.
\label{crit eight}
\eeq

Finally we note that if $\vev{f_{,i}}=0$, the criteria in the flat
case would simplify to
requiring $a_i$ and $b_i$ relatively real, but this possibility
is excluded because it gives $\tilde m_\lambda=0$, and
thus massless gauginos at tree level.

\subsection{The Polonyi Model}

Let us examine the implications of our criteria on a specific model.
The Polonyi model
\refmark{REFpolonyi}
is a simple SUGRA model, with a flat \Kpot, and only one
hidden field $z$ whose VEV breaks \susy.
Using this information, we write the quantities
defined in Section 4.1 as:

\bmulteq
& \vev{z} = b \, M, \ \vev{h} = m \, M^2 ,\ \vev{h'}  =  a^* \, m \, M  ,
  &\nonumber\\
& A = b^*(a+b) ,\
m_0 = e ^{{1\over 2}\abs{b}^2} m .
  &\eqn{onezVEVs}
\emulteq

\noi
In the Polonyi model, the hidden potential $h(z)$ has the specific form:
\beq
h(z) = m' \, M \, (z + \beta M) , \eqn{hidpot}
\eeq

\noi
where $\beta$ and $m'$ are in general {\it complex}.
The parameters $m$ and $m'$ are universally
defined as one
\refmark{REFbd,REFh,REFe,REFt,REFf,REFu,REFy,REFi},
$i.e$ $m' \equiv m$.  On the surface this seems silly, because
it is trivial to see that
$m = m'(b + \beta)$,
and $b$ and $\beta$ are both arbitrary complex numbers!  However
if one minimizes $V$ and uses the $\Lambda = 0$ condition $\abs{a+b}^2=3$,
one finds that $a$, $b$ and $\beta$ are necessarily {\it relatively} real,
and
$\abs{b + \beta} = 1$,
so that $\abs{m} = \abs{m'}$.  But this does {\it not} mean that
$m = m'$,
because they still differ by the phase of $\beta$.
Notice that since $a$ and $b$ are relatively real,
\pref{onezVEVs} implies that $A$ is manifestly real.
This is the basis of the claim
that the Polonyi model solves the SUSY \dn\ problem.
Unfortunately this satisfies only the first criterion in
\pref{crit three}.  The problem is that $m_0$ is not in general real,

\beq
\Arg m_0 = \Arg m = \Arg \beta m' ,  \eqn{arg m0}
\eeq

\noi
and neither is the product $m_0 \tilde m_\lambda$, whose phase we
can find using \pref{m0 mlambda}:

\beq
\Arg \left[ m_0 \tilde m_\lambda \right] =
\Arg \left[ \beta \vev{{\partial f \over \partial z}} \right].
\label{arg m0 mlambda}
\eeq

\noi
Both masses are invariant under a redefinition of $z$, so the phase
in \pref{arg m0 mlambda} cannot be rotated away.  If we can write
$f(z) = c_n z^n/M^n$, then one needs to have $\beta$ and the $c_n$ real.
We know of no mechanism to achieve this naturally.
As we said above, $\tilde m_\lambda =0$ ($\vev{\partial f/ \partial z} =0$)
solves the CPV problem, but gives an unacceptable mass spectrum.

We note that the Polonyi model has another naturalness problem coming from the
parameter $\beta$.
When we used the condition $\abs{a+b}^2=3$ to make
the cosmological constant $\Lambda$ vanish, we had no mechanism
to enforce it.
We had to arbitrarily {\it choose} $\abs\beta$ to be exactly $2-\sqrt3$
\refmark{REFt}.
In fairness, the cosmological constant is a problem
in all theories, and at least SUGRA models allow for $\Lambda=0$,
whereas global SUSY models do not \cite{REFax}.

\section{Concluding Remarks}

The SUSY contribution to the electric dipole
moment of the neutron (\dn) is quite large
unless one allows fine-tunings, or one has a large SUSY mass scale,
or one naturally sets the phases of the relevant parameters to zero.
Our criteria for the soft breaking parameters
employ the latter method. We showed that these criteria
can be written so that the MSSM gives no new contribution to \dn,
we investigated
the cases where the parameters were zero, and we derived forms for the
criteria in a large class of SUGRA models.
This allowed us to show that the Polonyi model does not naturally solve the
SUSY \dn\ problem.
We also showed that CKM induced finite loop effects in the
squark mass matrices give non-zero but negligible contributions to \dn,
and thus do not affect our conclusions.

We believe that $CP$ violating observables such as \dn\ should
be viewed as important phenomenological constraints, and that therefore
any serious model of \susy\ should
be consistent with the limit on \dn, and any model of SUSY breaking should
provide a solution to this
`SUSY \dn\ problem'.
For minimal \susyic\ models which give small to moderate superpartner masses
(of order the weak scale), our phases criteria should be looked upon as
tools for model builders for solving this SUSY \dn\ problem.
Non-minimal models will in general have more phases, so in most
cases our criteria will be a subset of the criteria in MSSM
extensions.

As we said, there is no new CPV in a
MSSM which satisfies our criteria.
Is this an acceptable situation? It would mean that \dn\ and
\de\ would be unobservably small.  This in and of itself is not
unacceptable---it would place SUSY in the same position as the SM---though
that may be disheartening for some experimentalists.  However, if a non-zero
electric dipole moment were observed in the near future, a MSSM
satisfying the criteria would be unable to account for it.
A complete understanding of the strong $CP$ problem would
be needed before conclusions could be drawn about \susy, but
it would be useful to know if a \susyic\ model satisfying our criteria
could explain an observable \dn\ without the need to appeal to
a small amount of strong CPV. A non-zero \de\ would demand such a mechanism.
There is also the possibility of generating the baryon asymmetry
of the universe at the electroweak scale \cite{BAU review},
which requires a source of
\CPV\ \cite{Sakharov}.
Some recent models of baryogenesis \cite{recent BAU} make use of
a moderate amount of CPV in the Higgs sector of a two doublet model,
which would not be present in a MSSM satisfying our criteria.

One might thus consider ways of generating
moderate amounts of CPV in models which have real tree level
MSSM parameters.  We explored such a mechanism,
which uses simple extensions to the MSSM to generate
$CP$ violating contributions to the Higgs potential
that are naturally suppressed by the size of loop effects \cite{modcp}.
This mechanism could generate \dn\ and \de\ near their experimental
bounds without resorting to fine-tunings or large mass scales,
and may be able to provide sufficient CPV for baryogenesis \cite{modcp}.
It may also be possible to induce moderate amounts of CPV in
SUGRA models through the spontaneous breaking
of horizontal symmetries \cite{Dine and Leigh}.
Also, an interesting  model of baryogenesis was recently proposed
\cite{Comelli et al}
using \CPV\
at finite temperature that might work with a MSSM satisfying
our criteria, though it is unclear that even the small
explicit phases they need could be supplied by radiative
corrections involving the CKM phase.
They might need a separate
mechanism for this  small amount of CPV, such as that of \cite{modcp}.
Of course the baryon asymmetry
could be generated at the GUT scale, and may have nothing to do with weak
scale $CP$ violation.

Moderate mass scales come out of most reasonable SUSY models.
To avoid the SUSY \dn\ fine-tuning problem, such models should
satisfy our criteria.  They would then tend to give a negligible
\dn\ and \de.
However,
such  models would not be immediately ruled out by the observation of
a non-zero electric dipole moment,
because there may be ways of naturally reintroducing a moderate amount of
\CPV\ into the theory.

\centerline{\Large\bf Acknowledgements}

I would very much like to thank Gordy Kane for suggesting this problem, and
for his many useful comments.
Thanks also to J. Soares and J. Wells.
This work was supported in part by a grant from the National Sciences
and Engineering Research Council of Canada.



\def\vol#1{#1}

\newpage

\centerline{\Large\bf Figure Captions}

{\bf Figure 1:} Gluino mediated contribution to the electric dipole moment
of the neutron.  The `X' indicates LR mixing of the down squarks is needed.

\

{\bf Figure 2:} Diagram which can give a finite contribution to
$\delta M_{LR}^2$ in \pref{MLR}.
Here the `X' indicates a mass insertion.
Note that the $H^+ \tilde U_L \tilde D_L$ vertex is unsuppressed \cite{REFg}.

\end{document}